\documentclass[twoside]{article}

\usepackage[latin1]{inputenc}

\usepackage[T1]{fontenc} 
\usepackage{amsmath,amsfonts,amssymb}

\usepackage{graphicx}
\usepackage{float}
\usepackage{epsf}
\usepackage{amsmath}
\usepackage{amssymb}
\usepackage{epsfig}
\usepackage{mathrsfs}
\usepackage{graphicx}

\usepackage{amssymb}
\usepackage{subfigure}
\usepackage{multirow}
\usepackage{color}
\usepackage{tabularx}

\usepackage[sc]{mathpazo} 
\usepackage[T1]{fontenc} 
\usepackage{microtype} 

\usepackage[hmarginratio=1:1,top=25mm,columnsep=20pt]{geometry} 
\usepackage{multicol} 
\usepackage[labelfont=bf]{caption}
\usepackage{booktabs} 
\usepackage{float} 
\usepackage{hyperref} 
\usepackage{authblk}

\usepackage{lettrine} 
\usepackage{paralist} 

\usepackage{abstract} 

\usepackage{fancyhdr} 
\title{\vspace{-15mm}\fontsize{24pt}{10pt}\selectfont\textbf{Polarimetry-based analysis of dipolar transitions of single colloidal CdSe/CdS dot-in-rods}} 

\author{
\large
C. Lethiec$^{1,2}$, F. Pisanello$^{3}$, L. Carbone$^{4}$, A. Bramati$^{5}$,L. Coolen$^{1,2}$, A. Ma\^itre$^{1,2}$
\thanks{maitre@insp.upmc.fr}\\ 
\normalsize $^1$ Sorbonne Universit\'es, UPMC Univ Paris 06, UMR 7588, INSP, F-75005, Paris, France \\ $^2$ CNRS, UMR 7588, INSP, F-75005, Paris, France \\  $^3$ Istituto Italiano di Tecnologia (IIT), Center for Bio-Molecular Nanotechnologies, Via Barsanti sn, 73010 Arnesano, Lecce, Italy\\ $^4$ NNL, Istituto Nanoscienze-CNR, c/o Distretto Tecnologico, Via per Arnesano 16, 73100 Lecce, Italy \\ $^5$ Laboratoire Kastler Brossel, Universit\'e Pierre et Marie Curie, Ecole Normale Supérieure et CNRS, UPMC Case 74, 4 place Jussieu, 75252 Paris Cedex 05, France.
\vspace{-5mm}
}

\date{}

\begin{document}

\maketitle 



\begin{abstract}
We prove experimentally, upon polarization analysis performed on a large statistic of single nanoemitters, that high quality core/shell CdSe/CdS dot-in-rods behave as linear dipoles. Moreover, the dipole in-plane and out-of-plane orientations could be assessed. We demonstrate in particular that, contrary to expectations, the emitting dipole is not aligned with the elongated axis of the dot-in-rod. Besides, the polarimetric measurements prove that the excitation transition cannot be approximated by a single linear dipole, contrary to the emission transition. Finally, we highlight that non-radiative channels of charge carrier recombination do not affect the dipolar nature of the radiative transitions.
\end{abstract}


\begin{multicols}{2} 

\section{Introduction}
The fast development of nanometer-sized single-photon sources realization \cite{Guyot,Reiss} has been acting as driving forces for the improvement of chemical approaches for the synthesis of high-quality and low-blinking colloidal quantum dots \cite{Wang09,Chen08,Spinicelli08}. Inorganic semiconductor nanocrystals developed according to a core/shell architecture, namely two different materials put together in a concentric topology and with a straddling alignment of the energy band gaps, display a very good fluorescence properties and high quantum efficiency at room temperature \cite{Kershaw}. In particular, colloidal core/shell nanocrystals in the rod shape made-up of a spherical dot of CdSe embedded into an anisotropic shell of CdS have been subject to many studies, especially as regards the dependence between their optical properties, such as lifetime \cite{Muller}, absorption anisotropy \cite{Kamal2012} and biexciton generation \cite{Pisanello13,Baer}, and the rod aspect ratio. Due to their elongated shape, they have the unique property of lying horizontally on the surface, which is an advantage when one wants to control the orientation of the elongated axis. A deterministic control of the orientation of the dipole is necessary to optimize the coupling of a semiconductor nanoemitter to photonic or plasmonic structures \cite{Belacel12}. In this respect the determination of the orientation of the emitting dipole associated to a dot-in-rod becomes crucial. Though one can expect that the dot-in-rod sets horizontally on the substrate, it is necessary to determine the influence of the elongated axis on the orientation of the dipole itself. Moreover, it has been proved that, due to a permanent density of surface-charges, the nanorod c-axis and therefore its internal electric polarization, can lie at a significant angle with respect to its sides \cite{Krauss04}. In this case, the induced electric field may modify the dipole orientation.\\
One usually distinguishes different types of emitting dipoles, depending on their nature: one can consider the case of a standard linear dipole, called "1D dipole". However, in many cases such as some molecules \cite{Pariser56}, some colored centers \cite{Epstein05} and some semiconductor quantum dots \cite{Lethiec13,Chung03,Brokmann04b,Vion09}, the emission originates from two degenerate states of orthogonal orientations. The emission corresponds then to an incoherent sum of two orthogonal 1D dipoles and is referred to as "2D dipole". The 1D or 2D nature of a given nano-emitter is related to the energy level structure and must be known before measuring the dipole orientation. \\
In this paper, we perform polarimetric measurements on CdSe/CdS dot-in-rods, whose chemical synthesis is described in references \cite{Carbone,Talapin}. Using the theoretical model developed in \cite{Lethiec13}, we demonstrate the linear nature and retrieve the orientation of the emitting dipole by analysing the emission polarization. We prove in particular that the dipole orientation is different from the orientation of the dot-in-rod elongated axis and we confirm these results on a large statistic of emitters with two different methods: polarization anisotropy measurements and defocused imaging technique. We demonstrate that the degree of linear polarization remains unchanged when the dot-in-rod is in an ionized state, proving that the polarization is not correlated to the local charge fluctuations. Finally, we perform a polarimetric study in excitation and highlight that the excited dipole is not fully polarized.\\
In the first section, we present the method of synthesis for the high-quality monodisperse emitters. Then we use the measurement of the degree of linear polarization to retrieve the in-plane and out-of-plane orientations of a single dipole. In a third part, we confirm the orientation on a larger statistics of dipoles by polarization anisotropy measurements and by defocused imaging. We then study how the intensity fluctuations affect the polarization of the dipolar transition. Finally, we perform a polarimetric analysis on the excited dipole.

\section{CdSe/CdS dot-in-rods synthesis}

\begin{figure}[H]
\begin{center}\includegraphics[width=7cm]{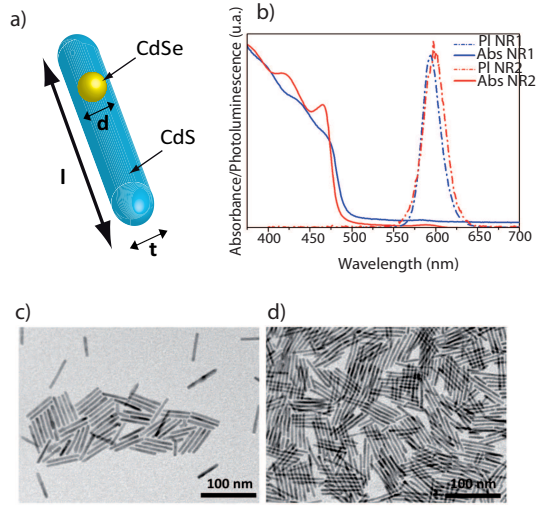}\end{center}
\caption{a) Sketch of the synthesized dot-in-rod, made up of a spherical core of CdSe embedded within the elongated shell of CdS. b) Absorbance (straight) and photoluminescence (dashed) spectra respectively of samples DR1 (blue) and DR2 (red). c) and d) Transmission electron microscopy (TEM) images of different sized CdSe/CdS dot-in-rods, respectively: c) sample DR1 (l=58 nm, d=2.7 nm, t=7 nm) and d) DR2 (l=72 nm, d=2.9 nm, t=4 nm).}
\label{Fig1}
\end{figure}

Two different specimens of CdSe/CdS dot-in-rods, hereafter respectively referred to as DR1 and DR2, are studied. The nanosized particles present characteristic core/shell morphology and, particularly in this instance, a round-shaped particle of CdSe is embedded within a rod-like shell of CdS, as illustrated in Fig.\ref{Fig1} a). The samples DR1 and DR2 have been synthesized following a seeded-growth approach described in \cite{Carbone} with a few modifications (see Supporting Information). The absorption and photoluminescence spectra associated to the DR1 and DR2 samples appear in Fig.\ref{Fig1} b). For both samples, the emission peak is centered at 600 nm.

Low-magnification transmission electron microscopy (TEM) analysis is performed on a Jeol JEM-1011 electron microscope operating at 100 kV, equipped with a CCD camera. TEM samples are prepared by drop-casting dilute nanocrystal solutions onto carbon coated copper grids. Diluted concentration of nanocrystal solution (< $10^{-8}$ M) significantly contributes to prevent the formation of bundles of vertically stacking dot-in-rods when drop-casted onto a substrate. The moderate particle concentration enlarges the interparticle distance and diminishes the likelihood of Van der Waals interactions between the long alkyl chains surrounding each nanocrystal. Additionally the nanocrystal length (> 50 nm) forwards a rod horizontal positioning than a vertical standing. We observe that the DR are horizontal and measure the geometric parameters l=58 nm, d=2.7 nm, t=7 nm and l=72 nm, d=2.9 nm, t=4 nm for DR1 and DR2 samples respectively. On this understanding we can expect that the dot-in-rods are lying horizontally on the substrate employed for optical observation (glass substrate of Root Mean Square roughness less than 1 nm as measured by Atomic Force Microscopy) as also evidenced in the TEM images (see Fig.\ref{Fig1} c) and d)).

\section{Polarization analysis on a single dot-in-rod}

\begin{figure}[H]
\begin{center}\includegraphics[width=7cm]{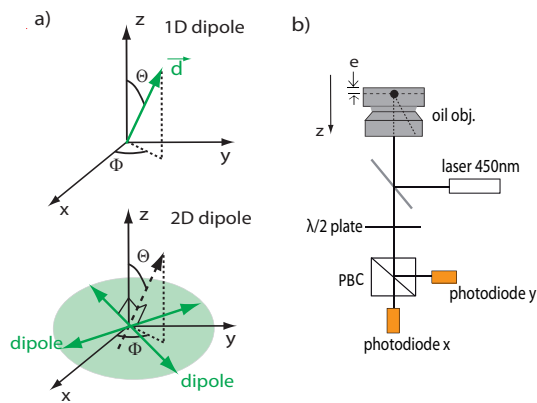}\end{center}
\caption{a) Orientation of a dipole in a (x,y,z) system. A 1D dipole sets along $\vec{d}$ oriented at ($\Theta$,$\Phi$) angles. For a 2D dipole, the angles ($\Theta$,$\Phi$) refer to the orientation of the normal to the dipoles plane, b) Schematic of the set up used to measure the dipole emitted intensity while rotating the half-wave plate. The emission is collected by an oil immersion objective. The z direction corresponds to the objective optical axis. A polarizing beam splitter cube separates the beam according to polarization whereas the signal is recorded by two photodiodes.}
\label{Fig1_1}
\end{figure}

In this section, we perform a polarimetric analysis on DR1 and DR2 CdSe/CdS dot-in-rods and we use the theoretical framework developed in \cite{Lethiec13} to retrieve the orientation ($\Theta$,$\Phi$) of a single nano-emitter, modelled as an emitting dipole.\\ 
The origin of the spherical system is chosen at the position of the dot and the z direction corresponds to the optical axis of light detection. In the case of a linear dipole, the orientation of the vector $\vec{d}$ is indexed by an out-of-plane angle $\Theta$ and an in-plane angle $\Phi$, as illustrated on Fig.\ref{Fig1_1} a) (above). For a 2D dipole (below), the angles ($\Theta$,$\Phi$) refer to the axis perpendicular to the plane formed by the two orthogonal dipoles.\\
The sample consists of dot-in-rods spin-coated on a glass coverslip and covered with 50 nm of PMMA, to protect the emitters from oxidation. The emitted signal is collected by an oil immersion objective of numerical aperture 1.4, mounted on an fluorescence confocal microscope. On Fig.\ref{Fig1_1} b), we represent the schematic of the set up. A half-wave plate is placed in the emission beam path, rotating with an angle $\alpha/2$. A polarizing beam splitter cube is located in front of the photodiodes, in order to separate the emission in x and y polarizations. The rotating half-wave plate and the polarizing beamsplitter cube placed one after the other act like a polarizer rotating the polarization analysis with an angle $\alpha$. This experimental configuration allows measuring both orthogonal polarizations on the two photodiodes. It is then possible to normalize the detected intensity on each photodiode by the total signal and avoid effects of the fluctuations of the total intensity.

\begin{figure}[H]
\begin{center}\includegraphics[width=8cm]{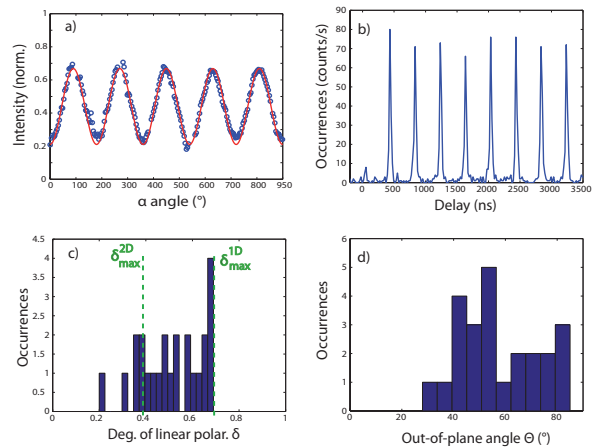}\end{center}
\caption{a) Circles: detected intensity as a function of polarization analysis angle $\alpha$ for a single DR2 dot-in-rod. The fitted curve (solid red line) corresponds to eq. (\ref{eq1}). The fitted value of $\delta$ for this particular dot-in-rod equals 52$\%$. b) Autocorrelation $g^{(2)}$ function obtained for the same dot-in-rod with a laser repetition rate 2.5 MHz. c) Histogram of experimental values of $\delta$ measured on 24 single DR2 dot-in-rods covered with 50 nm of PMMA. d) Histogram of experimental values of $\Theta$, deduced from Fig.\ref{Fig1_2} connecting $\delta$ to $\Theta$ for a 1D dipole, for the 24 single DR2 dot-in-rods: $\Theta$ ranges between 30$^\circ$ and 85$^\circ$.}
\label{Fig4}
\end{figure}

In Fig.\ref{Fig4} a), we plot as a function of the angle $\alpha$ the intensity emitted by a single DR2 dot-in-rod (blue circles), detected on the photodiode x, normalized by the total emitted intensity recorded by the two photodiodes. On each photodiode, the detected intensity is then oscillating between minimum $I_{min}$ and maximum value $I_{max}$ as the half-wave plate is rotating. In Fig.\ref{Fig4} b) is represented the autocorrelation curve corresponding to the same dot-in-rod than in a). Such a curve ($g^{(2)}$ function) represents the distribution of delays between a first photon on the photodiode 1 an the next photon on the photodiode 2. The zero photon coincidence on both photodiodes at zero delay (photon antibunching) proves that it is a single-photon source \cite{Brokmann04}, consequently proving that it corresponds to a single dot-in-rod \cite{Pisanello2010}.

As calculated in \cite{Lethiec13}, the detected intensity emitted by a 1D and a 2D dipole as a function of the rotating angle $\alpha$ can be theoretically written as the general expression\,:

\begin{equation}
I(\alpha) = I_{min} + (I_{max}-I_{min})\cos^2(\Phi' - \alpha)
\label{eq1}
\end{equation}

where $I_{min}$ and $I_{max}$ stand respectively for the minimum and maximum dipole detected intensity. In the case of a 1D dipole, $\Phi'$=$\Phi$ since the detected intensity is maximum when the axis of polarization $\alpha$ is aligned with the "bright-axis" of the dipole, referred to as $\Phi$. For a 2D dipole, $\Phi'$=$\Phi$+$\pi/2$. On the contrary, the detected intensity is minimum when the axis of polarization $\alpha$ equals to $\Phi$.\\
The phase of the measured modulation curve can be used to obtain $\Phi$, provided that the 1D or 2D nature is known (for instance, on Fig.\ref{Fig4} a), $\Phi'$=146$^\circ$). Moreover, the amplitude of this modulation, characterised by the degree of linear polarization $\delta$, allows determining the dipole out-of-plane orientation $\Theta$ \cite{Lethiec13}.\\
From the experimental curve, one usually extracts the factor $\delta$, defined as\,:

\begin{equation}
\delta = \frac{I_{max} - I_{min}}{I_{max} + I_{min}}
\label{eq3}
\end{equation} \\

\begin{figure}[H]
\begin{center}\includegraphics[width=7cm]{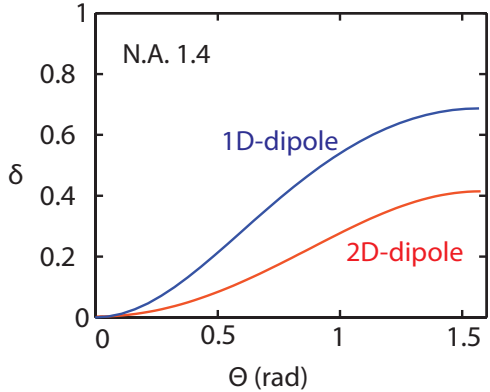}\end{center}
\caption{Calculated curves of the out-of-plane angle $\Theta$ as a function of the value of $\delta$ in both cases of a 1D (blue) and 2D dipole (red) located at 50 nm from a dielectric interface and observed with an oil objective of NA=1.4.}
\label{Fig1_2}
\end{figure}

In Fig.\ref{Fig1_2}, the simulated curves calculated for both 1D and 2D dipole provide the theoretical correspondence between $\delta$ and the out-of-plane angle $\Theta$ of the dipole. The model takes into account experimental values of numerical aperture NA=1.4, the glass and PMMA index 1.5 and a distance between the emitter and the PMMA/air interface e=50 nm (see Fig.\ref{Fig1_1} b)).

\begin{figure}[H]
\begin{center}\includegraphics[width=6cm]{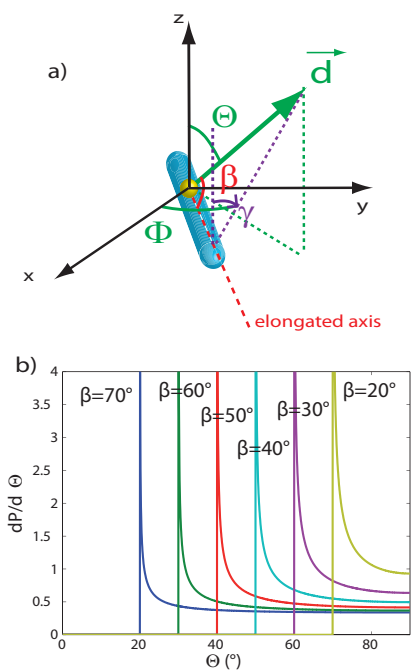}\end{center}
\caption{a) Orientation of the emitting dipole compared to the dot-in-rod. The angle between the dipole and the elongated axis is $\beta$. b) Calculated probability  density $\frac{dP}{d\Theta}$ for $\beta$ angles set from 20$^\circ$ to 70$^\circ$.}
\label{Fig4_1}
\end{figure}

As displayed in Fig.\ref{Fig1_2}, the correspondence between the degree of linear polarization $\delta$ and the out-of-plane angle $\Theta$ depends on the 1D or 2D nature of the dipole. The case $\Theta$=0$^\circ$ corresponds either to a vertical 1D dipole or to a horizontal 2D dipole (two horizontal dipoles), which in both cases presents a rotational symmetry configuration. Therefore, for $\Theta$=0$^\circ$, the degree of polarization $\delta$ is zero in both cases. It increases with $\Theta$ up to a maximum value which depends on the 1D or 2D nature of the dipole, the numerical aperture of the objective and the presence of a dielectric interface. The maximum value of the degree of linear polarization $\delta$, obtained for $\Theta=90^\circ$, is higher for a 1D dipole than for a 2D dipole (here respectively 70$\%$ and 40$\%$ for a dipole located at 50 nm from the interface and a NA=1.4), showing that the emission is less polarized for the 2D dipole as it is a sum of two incoherent dipoles. It is noteworthy that, in this configuration, even a perfect 1D dipole will not lead, for emission polarimetric analysis, to $\delta=100\%$. The bijective relation between $\Theta$ and $\delta$ allows retrieving the orientation of a given dipole of known 1D or 2D nature by measuring the degree of linear polarization $\delta$. Here we apply this method on dot-in-rods in order to retrieve both their nature and orientation.\\
In Fig.\ref{Fig4} a), we used the calculated relation (\ref{eq1}) between the intensity detected and the polarization analysis angle $\alpha$ to fit (in red) the experimental curve. The fit is very satisfactory, leading to $\delta$=0.52 with an uncertainty of 0.05. The same polarization measurements were performed on more than 20 nanorods, whose results are displayed in Fig.\ref{Fig4} c).\\
The histogram of experimental $\delta$ shows values mostly larger than 0.4 and up to 0.7. According to the simulated curves of $\delta$ as a function of $\Theta$ plotted in Fig.\ref{Fig1_2}, $\delta$ is always smaller than 0.4 in the case of a 2D dipole whereas it can reach 0.7 in the case of a 1D dipole. Then the values of $\delta$ measured on these 24 DR2 dot-in-rods highlight that these dot-in-rods are not 2D dipoles and can be analysed as 1D dipoles. Similar results have been reported on semiconductor nanorods \cite{Alivisatos01}. Moreover, the fine-structure model for semiconductor nanorods developed in \cite{Efros04} and pseudopotential calculations for CdSe rods \cite{Hu2002} predict that the emission polarization is almost 100$\%$ linear.\\
The experimental determination of $\delta$, using the theoretical curve (see Fig.\ref{Fig1_2}) calculated for a 1D dipole, leads to values of the out-of-plane angle $\Theta$ which range from 30$^{\circ}$ to 85$^{\circ}$. This indicates that most of the dipoles are not horizontal. However, as indicated by our TEM analysis as well as by AFM investigation performed on spin-coated nanorods \cite{Hadar2013}, such elongated dot-in-rods tend very likely to lie horizontally with the substrate. The measured values of $\Theta$ lead to the conclusion that the axis of the emitting dipole does not match the horizontal elongated axis of the dot-in-rod. Control experiments with shorter DR1 dot-in-rods with the same dot size have been performed. The resulting values of $\Theta$ were in the same range, included in [25$^\circ$-80$^\circ$]), proving that the dipoles of these DR1 dot-in-rods are not horizontal either. Because of this similar distribution of out-of-plane angles for both types of samples, we conclude that the geometrical size of the dot-in-rod is not the determining factor for the orientation of the dipole. This conclusion is consistent with the fact that the states involved in the emission are localized in the CdSe core which, in both cases, has the same size.\\
In Fig.\ref{Fig4_1} a), we represent the orientation of a 1D dipole associated to a dot-in-rod compared to the elongated axis of the dot-in-rod. The dipole has been located in the dot position. The angle between the dipole and the dot-in-rod is labelled as $\beta$. It must be taken into account that the plane defined by the dot-in-rod axis and the dipole axis is most likely tilted from the vertical plane, by an angle labeled $\gamma$ (see Fig.\ref{Fig4_1} a)) Therefore, the angle $\Theta$ equals $\pi/2-\beta$ only for $\gamma$=0 and $\Theta \le \pi/2-\beta$ in general.

In order to extract information on $\beta$ from the histogram of measured $\Theta$ angles in Fig.\ref{Fig4} d), one can establish the relation between the angles $\Theta$, $\gamma$ and $\beta$\,:

\begin{equation}
\Theta = \arccos (\sin\beta\cos\gamma)
\label{eq3}
\end{equation}

For a given $\beta$, assuming a uniform distribution of $\gamma$ between $-\pi/2$ and $\pi/2$ ($\frac{dP}{d\gamma}=\frac{1}{\pi}$), we find the theoretical distribution of $\Theta$\,:

\begin{equation}
\frac{dP}{d\Theta} = \frac{dP}{d\gamma}\frac{d\gamma}{d\Theta} = \frac{1}{\pi}\frac{\sin\Theta}{\sqrt{\sin^2\beta-\cos^2\Theta}}
\end{equation}

\begin{figure}[H]
\begin{center}\includegraphics[width=6cm]{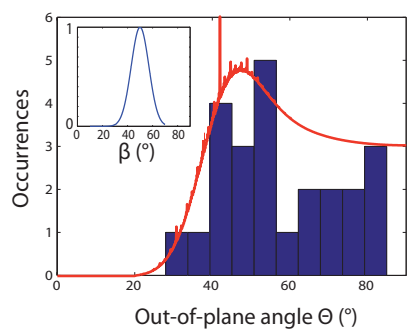}\end{center}
\caption{Histogram of experimental values of $\Theta$, as displayed in Fig.\ref{Fig4} d). The red curve stands for the simulated distribution of $\Theta$ corresponding to a gaussian distribution of $\beta$ centered in 50$^\circ$ with a FWHM of 20$^\circ$ (inset).}
\label{Fig12}
\end{figure}

In Fig.\ref{Fig4_1} b), we represent the probability density $\frac{dP}{d\Theta}$ of $\Theta$ for different values of $\beta$ from 20$^\circ$ to 70$^\circ$. A maximum is reached for each curve at $\Theta=\pi/2-\beta$, with no possible value of $\Theta$ below $\pi/2-\beta$ (as it is clear from Fig.\ref{Fig4_1} a)).\\
We find no occurrence for $\Theta$ angles lower than 30$^\circ$ and a peak between 30$^\circ$ and 50$^\circ$. As we did not measure any $\Theta$ angles lower than 30$^\circ$, we conclude that no dot-in-rods presents a value of $\beta$ higher than 60$^\circ$. Moreover, the fact that the occurrences of $\Theta$ are higher for $\Theta$ values between 30$^\circ$ and 60$^\circ$ is a good indication that most of the $\beta$ angles should take a range of values between 30$^\circ$ and 60$^\circ$.\\
In Fig.\ref{Fig12}, we overlay the histogram of measured $\Theta$ angles and the histogram (in red) of $\Theta$ values calculated from eq.(\ref{eq3}) assuming a gaussian distribution of angles $\beta$ centered in 50$^\circ$ with a FWHM of 20$^\circ$.    

The good correspondence observed in Fig.\ref{Fig12} between the measured and the simulated histograms of $\Theta$ confirms that the angles $\beta$ between the dipole and the dot-in-rod axis are most likely between 30$^\circ$ and 50$^\circ$.

\section{Polarization analysis on a larger statistics of dot-in-rods} 

\begin{figure}[H]
\begin{center}\includegraphics[width=8cm]{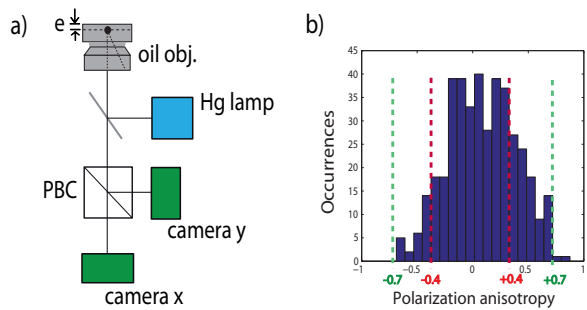}\end{center}
\caption{a) Schematic of the polarization anisotropy measurement set-up. The sample, with emitters deposited on a planar substrate and protected by a layer of PMMA, is excited with a Hg-lamp and observed with an immersion objective. A polarizing beam splitter cube is placed in front of two CCD-cameras in order to image  separately x and y polarization. b) Distribution of anisotropy measurements A extended to 413 DR2 dot-in-rods embedded into 50 nm thick film of PMMA.}
\label{Fig6}
\end{figure}

In this section, we confirm the previous results with a polarimetric analysis on a larger statistic of dot-in-rods. We use an alternative equivalent method, proposed by Chung et al. \cite{Chung03}, in order to confirm the linear nature of the dipoles associated to the dot-in-rods as well as their orientation. The set-up is presented in Fig.\ref{Fig6} a). It allows imaging simultaneously the x and y-polarized emission of the same emitters on two charge-coupled-device (CCD) cameras situated after a polarizing beam splitter cube. 
This polarizing beam splitter cube has been placed between the eyepiece and the CCD detector (onto which the optical beam is collimated in first approximation because the angle of the beam arriving on the CCD camera is divided by the high value of the lens magnification (x100)), so that the theoretical model of reference \cite{Lethiec13} can be used. It is then possible to obtain for each nanoemitter the experimental value of the polarization anisotropy A, defined as\,:

\begin{equation}
A=\frac{I_x - I_y}{I_x + I_y}
\end{equation} \\ 

with $I_x$ (resp. $I_y$) representing the value of the measured intensity of a given emitter on the CCD camera x (resp. y).

\begin{figure}[H]
\begin{center}\includegraphics[width=7cm]{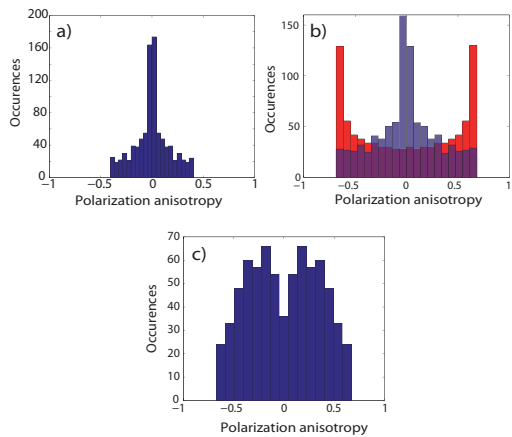}\end{center}
\caption{a) Calculated polarization anisotropy A for a collection of randomly-oriented 2D dipoles, emitting at $\lambda$=600 nm, overcoated by a layer of PMMA and thus located at 50 nm from the air interface. b) Calculated polarization anisotropy histograms (same experimental conditions than in a)) for a collection of 1D dipoles randomly-oriented (blue) and horizontally-lying ($\Theta=90^\circ$) onto the substrate (red). c) Calculated polarization anisotropy histograms for a collection of 1D dipoles with $\Theta$ distributed according to the histogram in Fig.\ref{Fig4} d).}
\label{Fig7}
\end{figure}

We measured the anisotropy polarization A on a collection of several hundreds of dot-in-rods. In Fig.\ref{Fig6} b), we plot the experimental histogram of polarization anisotropy for 413 DR2 dot-in-rods.\\
We know from the model presented in \cite{Lethiec13} that for any value of $\Theta$, it is not possible to measure A higher than 0.7 for a 1D dipole and 0.4 for a 2D dipole. Therefore, the experimental histogram of polarization anisotropy allows highlighting the 1D or 2D nature of a collection of dipoles. We simulate on Fig.\ref{Fig7} a) the distribution of A for a collection of 2D dipoles. If the dot-in-rods were 2D dipoles, we deduce than the associated histogram of polarization anisotropy would not reach values higher than 0.4. The fact that the experimental histogram displays A values higher than 0.4 and up to 0.7 implies that the dot-in-rods are 1D dipole for a good statistic. This result is consistent with the above reported polarization analysis. In Fig.\ref{Fig7} b), we plot both histograms for 1D dipoles corresponding respectively to an isotropic distribution of $\Theta$ (in blue) and to a collection of horizontal dipoles ($\Theta=90^\circ$) on the surface (in red). None of these two histograms match with the experimental histogram of Fig.\ref{Fig6} b. This consideration indicates that the dipoles orientations are neither horizontal nor completely isotropically distributed on the substrate. It is necessary then to go further in the discussion in order to simulate a reasonable representation of our experimental histogram.  

In order to refine the description of the orientation, we choose a theoretical distribution of $\Theta$ based on the histogram of the Fig.\ref{Fig4} d). On this figure, we note that $\Theta$ is distributed between 30$^\circ$ and 85$^\circ$, with more occurrences around 40$^\circ$ and 50$^\circ$. Therefore, we simulate in Fig.\ref{Fig7} c) a histogram of polarization anisotropy corresponding to 720 1D dipoles with $\Theta$ values fixed to 30$^\circ$ (90 dipoles), 40$^\circ$ (180 dipoles), 50$^\circ$ (180 dipoles), 60$^\circ$ (90 dipoles), 70$^\circ$ (90 dipoles) and 80$^\circ$ (90 dipoles), which represent a distribution of out-of-plane angles $\Theta$ close to that measured on Fig.\ref{Fig4} d). The calculated histogram of polarization corresponding to this distribution is reasonably similar to the experimental histogram of polarization anisotropy on Fig.\ref{Fig6} b). This result is in good agreement with the values of $\Theta$ we established in section 3 and confirms the result that the dipole corresponding to a dot-in-rod is not aligned with the geometrical dot-in-rod axis.

\section{Defocused imaging}

\begin{figure}[H]
\begin{center}\includegraphics[width=6cm]{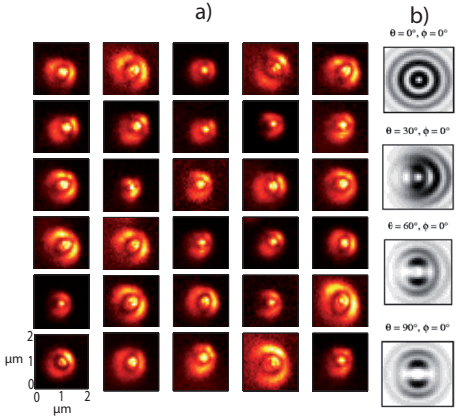}\end{center}
\caption{a) Panel of defocused images (area size: 2x2 $\mu$m$^2$) with $\delta$z=1 $\mu$m of 30 different DR1 dot-in-rods observed through an oil objective of NA=1.4. b) Theoretical defocused images for $\Theta$ respectively corresponding to 0$^\circ$, 30$^\circ$, 60$^\circ$ and 90$^\circ$ reported by Patra et al. in \cite{Patra04} considering an oil immersion objective of NA=1.2 at a distance $\delta$z=1.2 $\mu$m from the sample. The white colour stands for the lower intensity zones.}
\label{Fig8}
\end{figure}

In this section, we confirm, by defocused imaging technique, that the dipoles are not parallel with the substrate; this will help us to rule out the option of horizontal dot-in-rods with a non-1D dipole and prove that the values of $\delta$ lower than 0.7 cannot be justified by a heterogeneous population of 1D and 2D dipoles, but effectively by tilted linear dipoles. Defocused microscopy is a successful technique of imaging \cite{Pisanello2010,Jasny97,Brokmann05}, which allows probing the emission pattern of a dipole by moving the objective away from the sample of a distance as large as 1 $\mu$m. Each defocused image provides information about the orientation of a given dipole. In Fig.\ref{Fig8}, we plot the defocused images acquired on 30 DR1 dot-in-rods. It appears evident that only a few of these patterns meet the central symmetry that could be expected if the dipoles were lying horizontally. However, the quantitative derivation of the orientation angles is a delicate procedure. The precise retrieving of the orientation of a dipole close to an interface from a defocused image requires non-trivial calculations which must take into account the emission in near field to fit the images and properly extract quantitative pieces of information. We then compare the collected images with those simulated by Patra et al. \cite{Patra04}, which were calculated for similar observation conditions.

From that theoretical reference, central symmetry of the defocused pattern corresponds to a horizontal dipole ($\Theta$=90$^\circ$). On the opposite, when the dipole is standing up on the substrate ($\Theta$=0$^\circ$), the defocused image consists in concentric circles. In the intermediate situations where the dipole is tilted on the surface, the defocused image presents only one axis of symmetry. Our experimental defocused images display such a single axis of symmetry. We can thus qualitatively conclude that most of the dipoles are tilted on the surface. Moreover, none of them is standing up on the substrate (perfect concentric circles would be expected). These results corroborate the polarization analysis measurements reported above.

\section{Dipolar transition}

\begin{figure}[H]
\begin{center}\includegraphics[width=6cm]{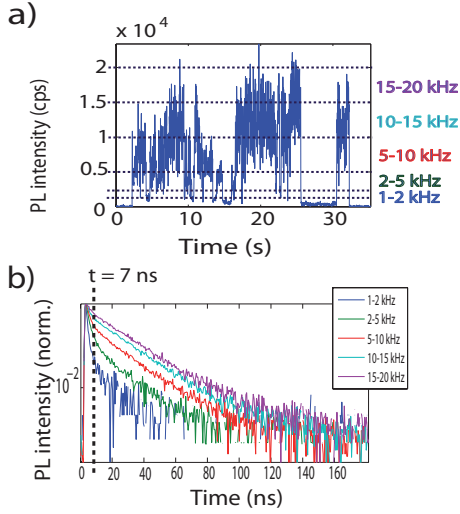}\end{center}
\caption{a) Intensity of a single DR2 dot-in-rod as a function of time. Photoluminescence intensity has been separated (dotted line) into different intensity levels of the time-trace to isolate bright and intermediate levels of emission, b) Emission decay curves analysed separately for the different ranges of intensity defined in a). The lifetime is higher for the bright state (15-20 MHz) than for intermediate levels because of the fast non-radiative transitions.}
\label{Fig2}
\end{figure}

We have proved above that the dipolar transition corresponding to the total emission of a CdSe/CdS dot-in-rod is 1D, in agreement with theoretical reports \cite{Efros04}, and that the dipole orientation is not the same as the rod long axis. However, it has been demonstrated that dot-in-rods luminescence fluctuates between dark, intermediate and bright states, giving rise to an overall emission instability \cite{Pisanello13}. Such events have been widely discussed in literature and recently assigned to different nanoparticle charged states \cite{Pisanello13,Galland11,Galland12}. The dark state is most likely due to a charging of the particle by trapping of a charge carrier in its shell. When the particle is charged, Auger-assisted non-radiative recombination (rate $\Gamma_{NR}$) competes with radiative recombination (rate $\Gamma_{R}$), leading to dark ($\Gamma_{NR}$ $\gg$ $\Gamma_{R}$) and intermediate ($\Gamma_{NR}$ $\sim$ $\Gamma_{R}$) states. In order to characterize and observe the intensity fluctuations of a single dot-in-rod, we plot in Fig.\ref{Fig2} a) the intensity time trace of one single DR2 dot-in-rod in the detection configuration of the Fig.\ref{Fig1_1} b), excited by a 450 nm pulsed laser. The emission exhibits different photoluminescence intensity levels corresponding to bright, intermediate and dark states. We divided the photoluminescence intensity in several intensity ranges, as displayed by the dotted lines, and plot in Fig.\ref{Fig2} b) the decay curves accumulated during the periods corresponding to the different intensity ranges defined in a). The lifetime $\frac{1}{(\Gamma_{R}+\Gamma_{NR})}$ is then different depending on the considered intensity range, as it has been observed on CdSe/ZnS nanocrystals in \cite{Fisher}.\\
For the bright state, the decay curve can be fitted by a mono-exponential curve with a well-defined lifetime (70 ns measured in Fig.\ref{Fig2} b) for intensities in the 15-20 MHz range). The presence of an additional fluctuating $\Gamma_{NR}$ rate in the intermediate states appears as a faster non-exponential decay.\\

\begin{figure}[H]
\begin{center}\includegraphics[width=6cm]{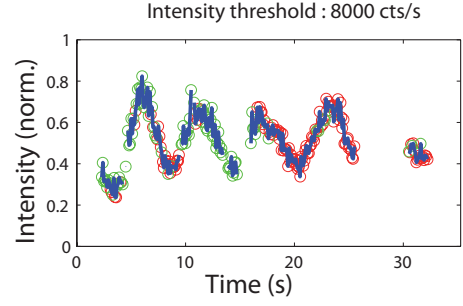}\end{center}
\caption{Detected normalized intensity while rotating the polarization of a single dot-in-rod emission as a function of time for an intensity of emission below and above the threshold $I_{th}=8000$ counts/s. The blue line corresponds to the total detected signal. The green (resp. red) circles stand for the photons which intensity is below (resp. above) the threshold.}
\label{Fig3}
\end{figure}

In order to determine if dipolar transitions with different orientations are involved in the different emission states of the dipole, we represent in Fig.\ref{Fig3} as a blue line the normalized intensity emitted by the single DR2 dot-in-rod studied in Fig.\ref{Fig2} a). The interruption of the curve corresponds to dark states, as it appears by comparing with the emission curve (in grey) reproduced from Fig.\ref{Fig2} a). A polarization degree $\delta=30\%$ is found. We introduce a threshold of intensity set to 8000 counts/s. Then we can consider separately the intermediate level (defined as intensity below 8000 counts/s) and the bright-state (emitted intensity above 8000 counts/s) and distinguish their response to polarization measurements. The red circles correspond to the signal recorded when the intensity is above the threshold. The green circles stand for the signal corresponding to the intensity below the threshold. 

In Fig.\ref{Fig3}, we note that the normalized intensity curves are superimposable in terms of amplitude and phase for the emitted signal below and above the threshold of intensity. The degree of linear polarization $\delta = \frac{I_{max} - I_{min}}{I_{max} + I_{min}}$ remains the same for any intensity on both sides of the threshold. This result indicates that the polarization of the energy level for the ionized emitter is the same as for the neutral emitter. This result allows us to conclude that the polarization of a dot-in-rod is not modified by the presence of non-radiative recombination channels.

\section{Polarization analysis of the excitation}

\begin{figure}[H]
\begin{center}\includegraphics[width=8cm]{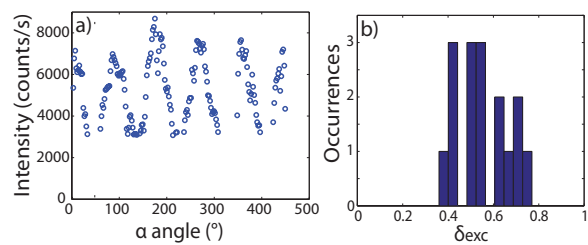}\end{center}
\caption{a) Circles: detected intensity emitted by a single DR1 dot-in-rod as a function of polarization excitation angle $\alpha$. $\delta_{exc}$ equals 0.49 for this single DR1 dot-in-rod, b) Histogram of experimental values of $\delta_{exc}$ for 16 DR1 dot-in-rods covered with PMMA.}
\label{Fig5}
\end{figure}

We consider in this section the result of polarimetric study of the excited dipole. In our set up, we assure the polarization of the 450 nm excitation laser diode by setting a polarizer in the beam. We then rotate the polarization with a half-wave plate placed in the excitation beam. The half-wave plate is continuously rotated with an angle $\alpha/2$ and the photoluminescence intensity is measured on both photodiodes. We excite a single dot-in-rod and we plot in Fig.\ref{Fig5} a) the emitted intensity as a function of the angle $\alpha$. In the same way as for emission, we define the degree of linear polarization of \textit{excitation} as $\delta_{exc}$=($I_{max}-I_{min})/(I_{max}+I_{min}$) and measure it on 16 single DR1 dot-in-rods. The signal, in this case of polarization analysis of the excitation, is more sensitive to the dot-in-rod total intensity fluctuations since it cannot be normalized. The experimental degrees of polarization $\delta_{exc}$ are summarized in the histogram of Fig.\ref{Fig5} b). \\

In Fig.\ref{Fig5} a), we measure $I_{min}\approx$ 2000 counts/s and $I_{max}\approx$8000 counts/s so the measured value of $\delta_{exc}$ is lower than unity when rotating the polarization of the excitation beam. According to the histogram of the experimental values of  $\delta_{exc}$ displayed in Fig.\ref{Fig5} b), $\delta_{exc}$ is included in [0.4;0.8] for the 16 studied DR1 dot-in-rods. 

Let us point out that the theoretical modelling and the interpretation of an excitation polarimetric curve are not the same as the ones for emission. The orientation of the exciting electric field $\vec{E}_{exc}$ at the position of the emitter is very close to the orientation $\vec{u}_{\alpha,\text{exc}}$ of the excitation polarizer (even when taking into account a high objective numerical aperture), as demonstrated in \cite{Ha99}. Therefore, for a 1D dipole $\vec{d}_{exc}$, one can write the intensity $I(\alpha_{exc})$ of the emission signal as \cite{Lethiec13,Ha99}\,:

\begin{equation}
\begin{split}
I(\alpha_{exc}) \propto  |\vec{d}.\vec{E}_{exc}|^2 \propto I_0|\vec{d}_{exc}.\vec{u}_{\alpha}|^2
\end{split}
\label{exc}
\end{equation}

So that, for an orientation ($\Theta_{exc}$,$\Phi_{exc}$) of the excited dipole, $I(\alpha_{exc}) \propto \cos^2(\Phi_{exc}-\alpha)\sin^2\Theta_{exc}$. It turns out that a $\delta_{exc}$ equal to unity is expected, for any out-of-plane orientation $\Theta_{exc}$. This is completely different from the case of an emitting dipole as explained in section 3 (the rotation of the excitation polarization is known to be not relevant to measure the full in-plane and out-of-plane orientation in the case of a linear dipole \cite{Guttler96}). The fact that we measure values of $\delta_{exc}$ much lower than unity in Fig.\ref{Fig5} b) indicates that the dot-in-rod is not equivalent for its excitation to a linear dipole, at least for the excitation wavelength used in this work.\\
The excitation degree of polarization has been found lower than unity also for some dot-in-rods \cite{Chen01} and semiconductor nanowires \cite{Lan,Shan01}. In reference \cite{Banin11}, the excitation polarization anisotropy of core/shell CdSe/CdS rod-in-rods at collective scale in solution has been proved to be lower when excited far above the band edge, as it is the case for our 450-nm excitation. This was explained by the structure of the higher excited energy levels which is a continuum of different polarization states. This consideration points out that the 1D or 2D nature and the orientation of the excited and the emitting dipoles is not equivalent as they correspond to different transitions.

\section*{Conclusion}

In this article, we report a thorough statistical investigation of the dipolar nature of core/shell CdSe/CdS elongated colloidal nanocrystals. We proved by polarization analysis based on a large ensemble of single CdSe/CdS dot-in-rods that both DR1 and DR2 dot-in-rods are linear emitting dipoles. By measuring the polarization anisotropy and using the defocused imaging technique, we confirmed the 1D nature of the dipoles that we proved by rotating the polarization of the emission and measuring the degree of linear polarization $\delta$, which turned out to be larger than 0.4. The axis of the linear emitting dipole has been found to be tilted from the geometric long axis of the dot-in-rod itself by an angle $\beta$. Although dot-in-rods spread out onto the substrate with their long axis parallel with the surface, we measured a dipole out-of-plane angle $\Theta$ between 40$^\circ$ and 60$^\circ$, so that the angle $\beta$ was estimated between 30$^\circ$ and 50$^\circ$. Besides, photons from bright and intermediate emitting states have been found to have the same polarization, letting us suggest that the transitions giving rise to photons emission remains the same even if the dot-in-rod is in a charged configuration. This last result proves that the fluctuations of the non-radiative decay do not change the polarity of the radiative transition. Finally, we showed that the excitation dipole of a dot-in-rod is not purely linear since the contrast is lower than unity in the experimental curves of the emitted intensity vs the rotating polarizer angle.\\
This study proves the potential of dot-in-rods as efficient single-photon polarized sources for quantum information.

\section*{Supporting Information}

\textit{Dot-in-rods synthesis.}\\
All synthesis steps are carried out under inert atmosphere of $N_2$.\\
\textit{Cd and S stock solution.} 0.060 g of CdO, 1.000 g of TOPO, 0.285 g of ODPA and 0.080 g of HPA are allowed to combine at 150$^\circ$C under vacuum. After almost one hour of dehareation and dehydration step the brown-coloured solution is exposed to a constant flow of N2 and heated to 320$^\circ$C until it becomes colourless and transparent. At this point, 1.5 g of TOP is injected and the heating supply definitely removed. When the sample reaches room temperature, a solution containing 0.120 g of S dissolved in 1.5 g of TOP is injected into the flask. At room temperature any chemical reaction between S and Cd is hampered. This colourless solution is finally stored under inert environment.\\
\textit{CdSe nanodots synthesis.} In brief, 3.700 g of TOPO, 0.280 g of ODPA and 0.060 g of CdO are stirred in a 50 mL flask, heated to 150$^\circ$C and exposed to vacuum for 1 hour. Afterwards, the solution is heated to 300$^\circ$C while flushing the flask with N2 until it turns transparent and colourless. At this stage, 1.5 g of TOP is injected into the reaction batch and the temperature set to 370$^\circ$C before injection of a Se-based solution (selenium powder 100 mesh 99.5$\%$ Sigma-Aldrich). After the temperature is stabilized to the set value, the Se-TOP solution (0.063 g Se + 0.575 g TOP) is swiftly injected and the growth is allowed to proceed for 1 min before definitely removing the heating mantle. CdSe nanodot solution is then transferred into a drybox and twice purified by sequential precipitation and re-solubilization with methanol and anhydrous chloroform, respectively. Finally the nanodots are dissolved in TOP in a final CdSe dots concentration of 7.5 $\mu$M. The as-described CdSe nanodots showed band-edge absorption peak at 536 nm.\\
\textit{CdSe/CdS DRs synthesis.}
In a typical synthesis step, 0.085 g of CdO (cadmium oxide 99.99$\%$ Sigma-Aldrich), 3.000 g of TOPO (trioctylphosphine oxide 99$\%$ STREM), 0.285 g of ODPA (octadecylphosphonic acid 99$\%$ Polycarbon Industries) and 0.080 g of HPA (hexylphosphonic acid 99$\%$ Polycarbon Industries) are combined together. The mixture is pumped to vacuum (50 mTorr) for one hour at 150$^\circ$C and then, after commuting to a N2 atmosphere, heated to 350$^\circ$C. At this point, 1.5 g of TOP (trioctylphosphine 99$\%$ STREM) is injected. After the temperature has newly re-set to 350$^\circ$C, a solution of S precursor-TOP-CdSe nanodots is swiftly injected into the flask. This solution is prepared by dissolving 0.120 g of S (sulfur 99.998$\%$ Sigma-Aldrich) in 1.5 g of TOP and adding 100 $\mu$l of a solution of TOP-dissolved CdSe dots (see above). After the injection, each CdSe/CdS nanocrystal solution is allowed to grow respectively for 8 min in the case of nanorods DR2 and 15 min for DR1. In the latter case, after 15 min of growth time, a Cd and S stock solution (see above) is dropwise injected (rate=300 $\mu$l/min, syringe volume=12 ml) at the fixed temperature for further 30 min . The synthesis reactions are stopped by removing the heating source. When the solution temperature cools down, the nanoparticles are transferred into a N2-supplied drybox and twice purified by precipitation with anhydrous methanol and re-solubilization in anhydrous chloroform. All samples are stored in the drybox until optical and morphological investigations.

\section*{Ackowledgements}

The authors acknowledge Francis Breton, Dominique Demaille and Willy de Marcillac for their technical support, Catherine Schwob, Paul B\'enalloul and Jean-Marc Frigerio, Stefano Vezzoli, Mathieu Manceau and Godefroy Lem\'enager for productive exchanges. The authors thank the Agence Nationale de la Recherche (P3N Delight and JCJC Ponimi) and the Centre de Comp\'etence NanoSciences Ile-de-France (C'Nano IdF, Sophopol, NanoPlasmAA) for funding this work.



\end{multicols}

\end{document}